\documentclass{article}
\newcommand{\bfr}{\begin{flushright}}
\newcommand{\efr}{\end{flushright}}
 
\begin{document}
\title{Can Virtual Cosmic Strings Shift the Hawking Temperature?
}
\author{Kiyoshi Shiraishi\\
Akita Junior College, Shimokitade-Sakura, Akita-shi, \\Akita 010,
Japan
}
\date{Europhys. Lett. {\bf 20} (1992) 483
}
\maketitle
\begin{abstract}
The amount of the Hawking radiation in Hartle-Hawking vacuum around
the Schwarzschild black hole is calculated in the presence of the cloud
of virtual-cosmic-string loops near the black hole. Recently, Coleman,
Preskill and Wilczek have discussed the black-hole temperature in such
situation. We show, in this letter, it is hardly possible to find any
deviation in the Hawking radiation even if the virtual cosmic strings
exist.
\end{abstract}

Recently, Coleman, Preskill and Wilczek have discussed \cite{1} the
role of virtual cosmic strings around black holes with screened charge
in the Higgs phase. The electric field induced by the charge of the
black hole (quantum electric hair) is present in the locus of the
virtual (thin) strings near the horizon membrane. Thus, we can measure
the effect of the screened charge of the black hole in the vicinity of
the horizon surface.

The authors of ref.~\cite{1} have also shown that the charge of the
system reduces the temperature of the black hole by the discussion on
the partition function.
It is worth studying the possibility that we can observe the quantum
effects of the screened charge of black holes at a spatially-distant
place.

In the present letter, we explicitly evaluate the expression for the
Hawking radiation of chaeged particles in the Hartle-Hawking vacuum
\cite{2} at spatial infinity, taking the virtual string effect into
account. The ``effective'' temperature will be read from the amount of
the Hawking radiation.

Now suppose we live in the Higgs (screened) phase of Abelian Higgs
model in which the Higgs has $U(1)$ charge $Ne$ $(N>1)$. We consider
the ``thin string'' case, for a definite discussion. (In $D$
dimensions, a topological defect in the model has $(D-3)$ dimensional
extension.) The ``motion'' of the virtual string as a topological
defect generates the following gauge configuration at spatial infinity
\cite{1}:
\begin{equation}
\int_0^{\beta\hbar}A_\tau d\tau\stackrel{r\rightarrow
\infty}{\longrightarrow}\frac{2\pi k}{Ne}\,,
\label{1}
\end{equation}
where $\beta^{-1}=(D-3) \hbar c/(4\pi r_g)$ is the Hawking temperature
$(k_B=1)$; $r_g$ is the Schwarzschild radius. The integer $k$ is the
wrapping number of the string loop around the hole. Therefore, the
(quantum-mechanical) physical quantity at a distance from the black
hole can seemingly be affected by the ``topological'' gauge
configuration induced by the quantum fluctuation of cosmic strings near
the black hole.

As a probe, we consider a conformally coupled scalar field with unit
$U(1)$ charge $e$.

The free energy {\it density} of the Hawking radiation of the scalar
particles in the Hartle-Hawking vacuum at spatial infinity is written
as follows in the presence of the external field (\ref{1}) \cite{3}:
\begin{equation}
f_k=\frac{2\hbar
c\Gamma(D/2)}{\pi^{D/2}(\beta\hbar c)^D}\sum_{n=1}^\infty\frac{\cos(2\pi
kn/N)}{n^D}\,.
\label{2}
\end{equation}

The partition function labeled by the wrapping number $k$ is given by
\begin{equation}
Z_k=Z_{rad.}\times Z_{string}\,,
\label{3}
\end{equation}
where $Z_{string}$ is the partition function of the cosmic string and
is approximately given by
\begin{equation}
Z_{string}\approx \exp\left\{-\kappa
|k|\frac{2\pi^{(D-1)/2}}{\Gamma((D-1)/2)} r_g^{D-2}\right\}\equiv
\exp(-\kappa |k| A)\,,
\label{4}
\end{equation}
in the thin string limit, or the string tension $\kappa$ is very large,
$Z_{rad.}$ is given by
\begin{equation}
Z_{rad.}=\exp(-\beta V f_k)\,,
\label{5}
\end{equation}
where $V=L^{D-1}$ is the volume of the system. In the present case,
since we must treat the quantity at a far distance from the black hole,
we must take $L\gg r_g$. If $L\sim r_g$, we must seriously consider the
effect of string dynamics as well as the curved space effect. We should
also note that the inclusion of the partition function of the black hole
is unnecessary in (\ref{3}), because that is independent of $k$ in the
Schwarzschild case and can be treated as a direct product on the final
result.

The partition function of the system with charge $Q$ is obtained by
the projection
\begin{equation}
Z_Q=\sum_{k=-\infty}^\infty\exp(-i2\pi Qk/Ne \hbar) Z_k\,.
\end{equation}

Using (\ref{2}), (\ref{4}), and (\ref{5}), we can get $Z_Q$. If
$V/(\beta\hbar c)^{D-1}\gg 1$, we can consistently
approximate $Z_Q$ as
\begin{eqnarray}
& &Z_Q\sim\exp\left\{\frac{2\Gamma(D/2)\zeta(D)
V}{\pi^{D/2}(\beta\hbar c)^{D-1}}\right\}\left\{1+2\cos\frac{2\pi
Q}{Ne
\hbar}e^{-\kappa A}\exp\left(-\frac{4\Gamma(D/2)
V X}{\pi^{D/2}(\beta\hbar c)^{D-1}}\right)\right\}\,,\nonumber \\
& &
\end{eqnarray}
where $X$ is
\begin{equation}
X=\sum_{n=1}^\infty\frac{\sin^2(\pi n/N)}{n^D}>0\,.
\end{equation}
For $D=4$, $X=(\pi^2/6)\{(N-1)^2/N^4\}$. Note also that $X=0$ if $N=1$.

For the massless particles, we get the expression for the energy
density:
\begin{eqnarray}
&
&\varepsilon=(D-1)\beta^{-1}\frac{\partial Z_Q}{\partial V}\nonumber
\\
& &\sim\frac{2\hbar c(D-1)\Gamma(D/2)\zeta(D)}{\pi^{D/2}(\beta\hbar
c)^{D}}\left\{1-\frac{8\cos\frac{2\pi Q}{Ne
\hbar}}{\zeta(D)}e^{-\kappa A}\exp\left(-\frac{4\Gamma(D/2)
V X}{\pi^{D/2}(\beta\hbar c)^{D-1}}\right)\right\}\,.\nonumber \\
& &
\label{9}
\end{eqnarray}

The amount of the energy density is modified by the virtual string
effect. The difference is, however, very small. (In ref.~\cite{3}, we
have unfortunately missed the last exponential factor in (\ref{9}).)

Since the temperature of the black hole (at the zero-loop level)
$\beta^{-1}$ is of the order of $r_g$, $V/(\beta\hbar c)^{D-1}\gg 1$
must hold. Therefore, the last factor in (\ref{9}) gives rise to
the very severe suppression. 

To summarise, we have found that the
Hawking radiation of charged particles suffers from very tiny correction
from the cloud of virtual cosmic strings and thus we can hardly
``detect'' the modification in the distance far from the
black hole. A naive expectation of the effect of the
``topological'' configuration (\ref{1}) failed us. Since the problem of
``detection'' in the curved space is subtle, we could not give a
definitive statement actually. We can safely conclude, at least, that
the black hole  temperature does not shift by the virtual string
effect, according to the usual interpretation of the Hartle-Hawking
vacuum ((unstable) equilibrium between black holes and radiation).

Of course, we may observe the quantum process in the vicinity of the
black hole horizon. The evaporation of the black hole may be affected
by the cloud of the virtual string loops. To study the effect, we must
investigate the string dynamics in the curved space-time more closely.
It is a fascinating problem in future.


\end{document}